# Low-loss directional coupler for the C, L and U bands based on subwavelength gratings

Jaime Vilas, Raquel Fernández De Cabo, Irene Olivares, David González-Andrade, Aitor V. Velasco and Antonio Dias-Ponte

*Abstract*—Directional couplers are ubiquitous components for power distribution in silicon photonics integrated circuits. Despite significant advances in their performance, architectures providing tailorable coupling ratios over increasingly broad bandwidths are still sought after. Compact footprint and low losses are also essential features for circuits comprising multiple coupling stages. In this work, we propose a compact directional coupler with arbitrary coupling ratio, based on dispersion engineering through subwavelength metamaterials. Low losses and a flat spectral response over a broad bandwidth are experimentally demonstrated for multiple coupling ratios between 0.08 and 1. Results show average excess losses below 0.24 dB and coupling ratio deviations below ±1 dB for a 170 nm bandwidth completely covering C, L and U telecommunication bands (1505 - 1675 nm).

## I. INTRODUCTION

SILICON-ON-INSULATOR (SOI) platforms provide compact and efficient photonics integrated circuits (PICs) while leveraging the maturity of microelectronics fabrication for reduced manufacturing costs [1]. Optical power distribution is a key functionality in almost any PIC, with a particular impact on the performance of optical filtering in wavelength division multiplexing (WDM) [2]. WDM filters are often composed of cascaded Mach-Zehnder interferometer (MZI) stages [3], each MZI stage comprising in turn two 2×2 power splitters and an optical delay line. The spectral response of the filters is tailored by judiciously designing the optical path lengths and the power splitter coupling ratios ($c_r$) of each MZI stage. As such, the overall filter performance is highly dependent on the performance of each individual power splitter, requiring minimal excess losses (*EL*), broad bandwidth and tailorable $c_r$ (i.e., adjustable during design). These stringent performance requirements on power splitters also extend to other high-impact applications such as optical phased arrays [4] or on-chip spectrometers [5].

Directional couplers (DCs) are one of the most widely used 2×2 power splitters, owing to their structural simplicity and straightforward design [6]. DCs typically present low losses, and their $c_r$ can be readily selected through coupling length design. However, conventional DCs are hindered by limited bandwidth and fabrication tolerances. Several routes have been proposed to address these shortcomings, including curved DCs [7], tapered DCs [8] and adiabatic couplers [9]. Subwavelength gratings (SWG) have also proven to be a powerful tool for overcoming DC bandwidth limitations [10]–[14]. SWG consist of periodic arrangements of core and cladding materials, with a pitch ($\Lambda$) significantly smaller than the operational wavelength ($\lambda$). This configuration suppresses diffractive effects and behaves as a metamaterial that enables dispersion and refractive index engineering through geometrical design [15]–[17]. SWG-based DCs have been successfully designed and fabricated before, achieving unprecedented performance milestones. However, further performance improvements are still sought after.

In this Letter, we report on the design and experimental demonstration of a compact SWG-based DC, geometrically engineered to achieve ultra-low losses and a broad bandwidth with a tailorable $c_r$. Experimental measurements show average excess losses below 0.24 dB and coupling ratio deviations below ±1 dB over a 170 nm bandwidth, covering the C, L and U communication bands.

## II. PRINCIPLE OF OPERATION

The architecture of our 2×2 SWG DC is depicted in Fig. 1, highlighting its three sections and design variables. The symmetric coupling region, labelled A, consists of two parallel mono-mode waveguides with the same width and length ($L_c$), separated a distance $W_{sep}$, where input light is progressively coupled from one waveguide to another.

In order to minimize the spectral dependence of this coupling effect, a tailored SWG region is included both between the waveguides and on their external sides, reaching a total width of $W_{SWG}$. The coupling section hence comprises $N_p$ periods of constant pitch and fill factor ($ff = a/(a+b)$), with $a$ and $b$ being

This work is funded by the Spanish Ministry of Science and Innovation (MCIN/AEI/10.13039/501100011033) and the European Union "NextGenerationEU/PRTR" through grant DIN2020-011488 and Horizon Europe under project 190113917; by the MCIN/AEI/10.13039/501100011033 under grants PTQ2021-011974 and PID2020-115353RA-I00; European Social Fund Plus grant (PRE2021-096954); Community of Madrid – FEDER funds (S2018/NMT-4326, APOYO-JOVENES-KXHJ8C-16-VCKM78). (*Corresponding author: Jaime Vilas*)

Jaime Vilas, Irene Olivares, and Antonio Dias-Ponte are with Alcyon Photonics S.L, Madrid 28003, Spain (e-mail: j.vilas@alcyonphotonics.com, irene@alcyonphotonics.com, antonio.dias@alcyonphotonics.com).

Raquel Fernández de Cabo, and Aitor V. Velasco are with the Instituto de Óptica, Consejo Superior de Investigaciones Científicas (CSIC), Madrid, 28006, Spain (e-mail: r.fernandez@csic.es, a.villafranca@csic.es).

David González-Andrade is with Centre de Nanosciences et de Nanotechnologies, CNRS, Université Paris-Saclay, Palaiseau 91120, France (e-mail: david.gonzalez-andrade@c2n.upsaclay.fr).

Color versions of one or more of the figures in this article are available online at http://ieeexplore.ieee.org

the lengths of the core and cladding segments,

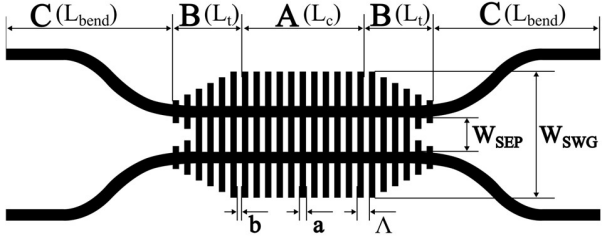

**Fig. 1.** Schematic of the proposed SWG DC.

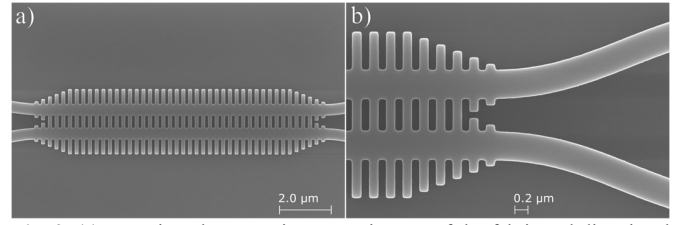

**Fig. 2.** (a) Scanning electron microscope images of the fabricated directional coupler with 32 SWG periods in the coupling region. (b) Close-up of the taper region.

respectively. Once SWG parameters are optimized for maximal bandwidth, $c_r$ can be readily tailored by selecting $N_p$, circumventing the need of numerical re-optimization of waveguide geometry of asymmetrical DCs [11] or MMIs [18], as well as enabling implementation of low $c_r$ without performance degradation.

The taper region (B in Fig. 1) features input and output SWG tapers engineered to provide a smooth transition of the optical mode between the strip waveguides and the SWG region, and hence reduce losses compared to direct interfaces [10],[14]. SWG tapers maintain the pitch and fill factor of the coupling region, with each taper comprising $N_t$ periods, resulting in a taper length of $L_t = N_t\Lambda$. Notice that both tapers present a slight power coupling between waveguides that needs to be taken into account in the overall design. A smaller $N_t$ hence implies lesser coupling at the tapers but needs to be balanced with the adiabatic transition condition for minimal losses.

Finally, the input/output regions, labelled C in Fig. 1, comprises S-bends that modify the distance between waveguides from/to a separation with negligible power coupling. The radius of the S-bend curves is optimized to minimize coupling in the interface between sections B and C, without inducing curvature losses.

### III. DESIGN

In order to optimize our SWG DC and evaluate its performance, we define the following figures of merit. Coupling ratio deviation ($\Delta c_r$) is defined as the deviation from the central $c_r$ within a given bandwidth. Said coupling ratio is computed as $c_r = P_{cross}/(P_{cross}+P_{through})$, where $P_{cross}$ and $P_{through}$ are the power output at the cross and through ports, respectively. Likewise, excess losses are defined by $EL = -10log(P_{cross}/(P_{cross}+P_{through}))$.

Since our SWG DC can be designed with an arbitrary $c_r$, we further define the mean excess losses ($\overline{EL}$) as the average $EL$ of all DCs under analysis with $c_r \in (0,1)$. We also define $EL_{0.5}$ as the $EL$ of a DC with $c_r = 0.5$, in order to facilitate comparison with other 3-dB DCs of the state of the art.

The proposed architecture was optimized for transverse electric (TE) polarization in an SOI platform with a Si core thickness of 0.22 μm, and SiO$_2$ cladding and buried oxide (upper cladding thickness of 2.2 μm). Wire waveguide width was set to 0.45 μm for mono-mode operation. Through full three-dimensional finite-difference time-domain (3D FDTD) simulations, we firstly optimized the geometric parameters of the coupling region. Notice that as $\Lambda$ increases, the beat length decreases [11], resulting in more compact devices at the expense of a higher sensitivity of $c_r$ to variations in $N_p$. This effect can be balanced by an increase in $W_{sep}$, which reduces coupling intensity and hence extends beat length. For a $W_{SWG}$ of 2.6 μm, EL and bandwidth optimization finally resulted in $\Lambda$ = 0.26 μm and $W_{sep}$ = 0.5 μm. We considered three $ff$ scenarios (i.e., 0.54, 0.5 and 0.46), resulting in minimum feature sizes (MFS) of at least 0.12 μm. These $ff$ variations also provide a tool to evaluate the effect of under/over-etching by 10 nm steps. The optimized SWG structure not only maximizes operational bandwidth while reducing losses and footprint, but also provides a more relaxed MFS compared to other state-of-the-art alternatives [10,11,14], hence facilitating fabrication through standard photolithography processes.

Then, taper length was optimized to 1.56 μm ($N_t = 6$). Taper impact on $c_r$ was evaluated by simulating a device with back-to-back tapers ($N_p = 0$), resulting in $c_r = 0.08$. This value sets the minimum design threshold for $c_r$ and needs to be taken into account when setting the coupling section length for a target $c_r$. Total coupling in the cross port ($c_r = 1$) is obtained for $N_p = 40$ at $ff = 0.54$. These limits result in device lengths (A+B sections) between 3 and 13.4 μm. Finally, we set the length of the S-bends ($L_{bend}$) to 5 μm, with a final separation of 2.95 μm between ports. We embedded the S-bends 0.65 μm into the tapers to further reduce coupling in the interface between sections B and C.

In order to experimentally demonstrate the performance of the device for different $c_r$, we fabricated 11 DCs with linearly increasing $N_p$ between 0 and 40, and replicated the designs for all three $ff$ scenarios. Fabrication was carried out using the electron-beam (e-beam) lithography process of Applied Nanotools (NanoSOI). The mask pattern was defined by exposing the resist to a 100 keV e-beam lithography system, followed by an anisotropic reactive ion etching process for pattern transfer. SiO$_2$ cladding was deposited by chemical vapor deposition. High-efficiency broadband SWG edge couplers were included to minimize fiber-chip coupling losses [18]. Several groups of reference waveguides without the DC structures were introduced, enabling output normalization. Fig. 2 presents scanning electron microscope images of the fabricated device, showing a slight over-etch which further accounts for the best performance being found for $ff = 0.54$.

### IV. RESULTS

Chip characterization was carried out using a tunable laser with a wavelength range between 1500 nm and 1680 nm. Input signal polarization was adjusted through a three-paddle fibre polarizer, followed by a linear polarizer, a half-wave plate and a lensed polarization-maintaining fibre. A free space polarimeter was used to verify polarization across the measured wavelength range. A Glan-Thompson polarizer at the chip output ensured the polarization state was correct, and a 40× microscope objective was used to focus the signal on a germanium photodetector.

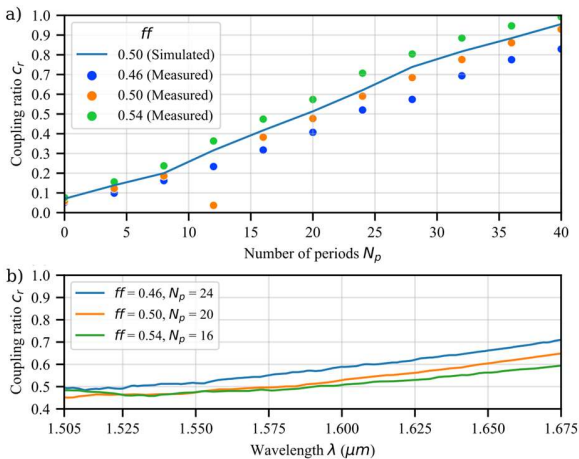

**Fig. 3.** (a) Normalized coupling ratio as a function of the number of periods of the coupling region for different fill factors, for a wavelength of 1.55 μm. (b) Spectral dependence of the coupling ratio = 0.5 devices.

Fig. 3(a) presents the measured $c_r$ at a wavelength of 1550 nm for all cases under analysis, showing very good correspondence with simulations and the typical sinusoidal dependence of $c_r$ with the coupler length. All 30 devices under analysis show consistent behaviour, with only one visible outlier ($ff = 0.5$, $N_p = 12$) likely caused by a faulty coupling structure. These results show straightforward $c_r$ configurability between 0.08 and 1 through unitary increments of $N_p$, in steps of 0.03 near $c_r = 0.5$, and in steps of 0.006 near $c_r = 0$ and $c_r = 1$. We can also observe that, for $N_p = 20$, an under/over-etching of 10 nm results in a $c_r$ variation of +0.09/-0.07, respectively. As shown in Fig. 3(b), underetching also implies a slight shift of the maximally flat region to higher wavelengths. These tolerance results are consistent with corner analysis simulations: for waveguide width variations of ±10 nm and thickness variations of ±5 nm, $\Delta c_r$ variations result in a worst-scenario bandwidth reduction of 25 nm.

Focusing on $ff = 0.54$, Fig. 4(a) presents the spectral response of the coupling ratio for all $N_p$ values under analysis. C, L and U communication bands are shaded within a total wavelength range of 170 nm. All examined DCs present a steeper slope at higher wavelengths, but still maintain $\Delta c_r < \pm 1$ dB across the full range. Individualized behaviour for the C, L and U bands is shown in Fig. 4(b), Fig. 4(c) and Fig. 4(d), respectively, demonstrating $\Delta c_r$ lower than ±0.2 (±0.3) dB for the C band, and lower than ±0.4 (±0.5 dB) for the L and U bands. These results highlight the versatility of the device for both broadband operation, and single band operation in case the application requires more stringent $c_r$ conditions.

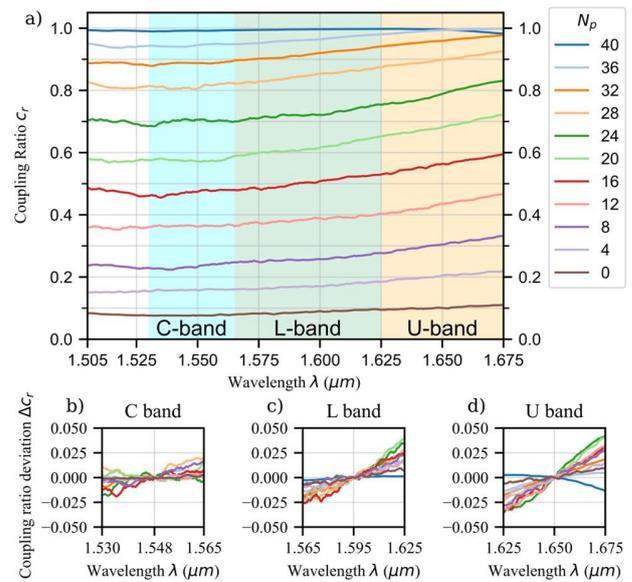

**Fig. 4.** (a) Coupling ratio for devices with filling factor = 0.54 as a function of wavelength for several coupling section lengths. Details on coupling ratio deviations for (b) C, (c) L and (d) U communication bands.

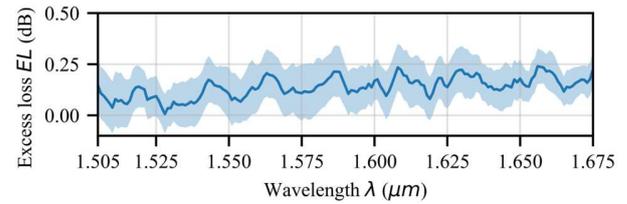

**Fig. 5.** Average excess losses for all coupling ratios under analysis (solid line) and standard deviation (shaded).

Finally, Fig. 5 plots $\overline{EL}$ for all DCs under analysis (i.e., average losses for $N_p$ from 0 to 40), as well as their standard deviation as a function of wavelength. $\overline{EL}$ below 0.24 dB are achieved for the full 170 nm band, with a reduction down to 0.2 and 0.23 dB in the individual C and L bands, respectively. $\overline{EL}$ standard deviation is below 0.14 dB in the full 170 nm range. Notice that measured losses fall close to the resolution limit of the measurement setup and, given the low-loss nature of the SWG structures, do not show any clear proportionality to $N_p$. In order to provide comparison with other 3-dB DCs in the state of the art, we also individualized the performance of our SWG DC with $N_p = 16$ ($c_r \approx 0.50$ in the measured bandwidth). Said DC exhibits very low excess losses of $EL_{0.5} < -0.28$ dB over the 170 nm band, which are reduced in the C and L bands down to -0.20 dB and -0.18 dB, respectively.

As shown in Table I, the proposed device compares favourably against other state-of-the-art SWG DCs [10-14], as well as to other splitter technologies [7,18]. Notice that bandwidths are defined by each author with different $EL$ thresholds, with the current work presenting the most stringent threshold. In the case of DCs with multiple $c_r$, performance values at $c_r = 0.5$ are displayed. $EL_{0.5}$ of our device are almost halved in comparison to the next best performing DC (in terms of excess losses) [12], while achieving nearly four times the bandwidth. Meanwhile, only the DC fabricated by Ye *et al.* [11] features a slightly broader bandwidth than that of the devices presented in this work, although defined at a losses threshold nearly four times higher.

TABLE I
PERFORMANCE COMPARISON BETWEEN STATE-OF-THE-ART 2×2 DIRECTIONAL COUPLERS.

| Ref | Structure | Coupling Length (μm) | Minimum Feature Size (nm) | $EL_{0.5}$ (dB) | Bandwidth (nm)[a] | Fully covered bands | Coupling ratios |
|---|---|---|---|---|---|---|---|
| [7] | Curved DC with straight sections | 20 | 200 | -[b] | 88 | C | 0.5 |
| [10] | Symmetric SWG DC | 19.2 | 61 | -[b] | 96[c] | C | 0.5 |
| [11] | Asymmetric SWG DC | 5.25 | 110 | 1 | 200 | S,C,L | 0.5 |
| [12] | Polarization independent SWG DC | 5 | 100 | 0.50 | 45 | C | 1 |
| [13] | Polarization independent SWG DC | 3.75 | 44[b] | 1 | 60 | C | 0.5 |
| [14] | Symmetric SWG DC | <14 | 80 | <0.70[b] | 100 | C | 0.5 − 0.8 |
| [17] | SWG MMI | 14 | 95 | 1 | >300 nm | S,C,L,U | 0.5 |
| This work | Symmetric SWG DC | 7.16 | 120 | 0.28 | 170 | C,L,U | 0.08 − 1 |

[a]Notice that bandwidth for each device is defined by their respective authors according with different $EL$ thresholds, as indicated in the previous column. [b]Extracted from figures or not explicitly stated. [c]Simulation results.

Finally, for the example of $N_p = 16$, our device has a total footprint of 3.4 × 16.38 μm$^2$, including the input/output S-bends; and of only 2.6 × 7.16 μm$^2$ if we only consider the taper and coupling regions. This makes our DC also competitive in terms of compactness.

## V. CONCLUSION

In conclusion, we have designed and fabricated a novel low-loss directional coupler based on subwavelength gratings metamaterials. Experimental characterization of the device for coupling ratios ranging from 0.08 to 1 shows average losses below 0.24 dB and maximum coupling ratio deviation below ± 1 dB across a bandwidth of 170 nm. Maximum coupling ratio deviations within the individual C, L and U bands are reduced to ±0.3 dB, ±0.5 dB and ±0.5 dB, respectively. Given the slightly steeper coupling ratio slope at higher wavelengths, device performance could be further improved by optimizing SWG metamaterial design for a greater central wavelength. Coupler length ranges from 3 μm ($c_r = 0.08$) to 13.40 μm ($c_r = 1$), including input/output tapers. These results showcase that the proposed design provides a tailorable coupling ratio while maintaining high performance and ensuring both a compact footprint-length and a manufacture-friendly minimum feature size. The performance of our device makes it as a strong candidate for general power distribution in SOI PICs, and in particular, for the implementation of multi-stage MZI-based optical filters.